\documentclass[aps,prl,floatfix,amsmath,amssymb,twocolumn]{revtex4-1}
\usepackage{amsmath, amssymb, graphics, mathrsfs, bm, stackrel,bbold}
\usepackage{subfigure}%http://texblog.wordpress.com/2007/08/28/placing-figurestables-side-by-side-subfigure/
\usepackage[usenames,dvipsnames]{xcolor}%http://en.wikibooks.org/wiki/LaTeX/Colors
\definecolor{Highlight}{rgb}{1,1,0.75}

\allowdisplaybreaks[1]
\usepackage{graphicx}
\usepackage[bookmarks={false}, pdfauthor={Rudro Rana Biswas}, pdftitle={Title}]{hyperref}%pdftex?
\hypersetup{colorlinks=true, linkcolor=BrickRed, citecolor=Violet, filecolor=OliveGreen, urlcolor=RoyalBlue, filebordercolor={.8 .8 1}, urlbordercolor={.8 .8 0}}%http://en.wikibooks.org/wiki/LaTeX/Hyperlinks
\usepackage[all]{hypcap}
\usepackage[export]{adjustbox}%extends includegraphics options, etc

\newcommand\ba{\begin{array}}
\newcommand\ea{\end{array}}
\newcommand\nn{\nonumber}
\newcommand\ri{\right}
\renewcommand\le{\left}

\newcommand{\feyn}[1]{#1\kern-0.45em/}

\renewcommand\a{\alpha}%metric symbol

\newcommand\mbA{\mbs{A}}
\renewcommand\b{\beta}%metric symbol
\newcommand\mbB{\mbs{B}}
\renewcommand\c{\psi}%accent

\renewcommand\d{\delta}%accent

\newcommand\e{\epsilon}

\newcommand\f{\phi}

\newcommand\p{\pi}
\newcommand\mbp{\mbs{p}}
\newcommand\mbpp{\mbs{\p}}

\newcommand\mbr{\mbs{r}}
\newcommand\mbR{\mbs{R}}

\newcommand\x{\xi}

\newcommand\mbz{\mbs{z}}

\newcommand\la{\langle}
\newcommand\ra{\rangle}
\newcommand\pd{\partial}
\newcommand\mc{\mathcal}
\newcommand\mb{\mathbb}
\newcommand\mbs{\boldsymbol}

\begin{document}
\title{Gauge-Invariant Variables Reveal the Quantum Geometry of \\Fractional Quantum Hall States}
\author{YingKang Chen}
%\email{rrbiswas@xx.edu}
\affiliation{Department of Physics and Astronomy, Purdue University, West Lafayette, IN 47907.}
\author{Rudro R. Biswas}
\email{rrbiswas@purdue.edu}
\affiliation{Department of Physics and Astronomy, Purdue University, West Lafayette, IN 47907.}

%\date{\today}
\begin{abstract}
Herein, we introduce the framework of gauge invariant variables to describe fractional quantum Hall (FQH) states, and prove that the wavefunction can always be represented by a unique holomorphic multi-variable complex function. As a special case, within the lowest Landau level, this function reduces to the well-known holomorphic coordinate representation of wavefunctions in the symmetric gauge. Using this framework, we derive an analytic guiding center Schr\"odinger's equation governing FQH states; it has a novel structure. We show how the electronic interaction is parametrized by generalized pseudopotentials, which depend on the Landau level occupancy pattern; they reduce to the Haldane pseudopotentials when only one Landau level is considered. Our formulation is apt for incorporating a new combination of techniques, from symmetric functions, Galois theory and complex analysis, to accurately predict the physics of FQH states using first principles.
\end{abstract}
%\pacs{00.00xx,yy}
\maketitle

Since the discovery of the quantum Hall effect \cite{1980-klitzing-uv,1982-tsui-jq}, substantial progress has been made in our understanding of topological quantum phases of matter, of which quantum Hall states are prototypical examples \cite{1981-laughlin-yq,1982-halperin-oq,1983-laughlin-fk,1984-pruisken-uq,1990-prange-xr,1995-wen-fk,1999-stormer-hl,2002-yoshioka-zl,2003-murthy-fk}. However, current theories have limited scope in reliably predicting the existence and properties of these states in experimentally achievable scenarios \cite{2017-hansson-oq,2007-jain-fk,2011-goerbig-fk}. In turn, such predictions are critical for establishing the viability and performance of topological quantum computation setups \cite{2008-nayak-fk}.

Specifically, there is a critical need for developing a universal theoretical approach for deriving the properties of strongly correlated topological quantum states, starting from realistic, experimentally-relevant microscopic Hamiltonians. Addressing this need, herein we establish a universal approach to formulating the quantum problem of calculating the properties of fractional quantum Hall (FQH) states at any filling fraction, involving any Landau level, by using the language of gauge-invariant variables (GIVs). The GIV representation \cite{1949-johnson-nr,2013-biswas-wr,2018-chen-zi} exploits the characteristic quantum geometry of topological phases \cite{2010-loring-jk} and allows us to isolate the essential one-dimensional physics of electronic motion inside a Landau level \cite{2005-bergholtz-qe,2005-abanov-fj,2007-horsdal-yw}.

%%%%%%%%%%%%%%%%%%%%%%begin%%%%%%%%%%%%%%%%%%%%%%
\begin{figure}[t]
\begin{center}
\resizebox{0.48\textwidth}{!}{\includegraphics{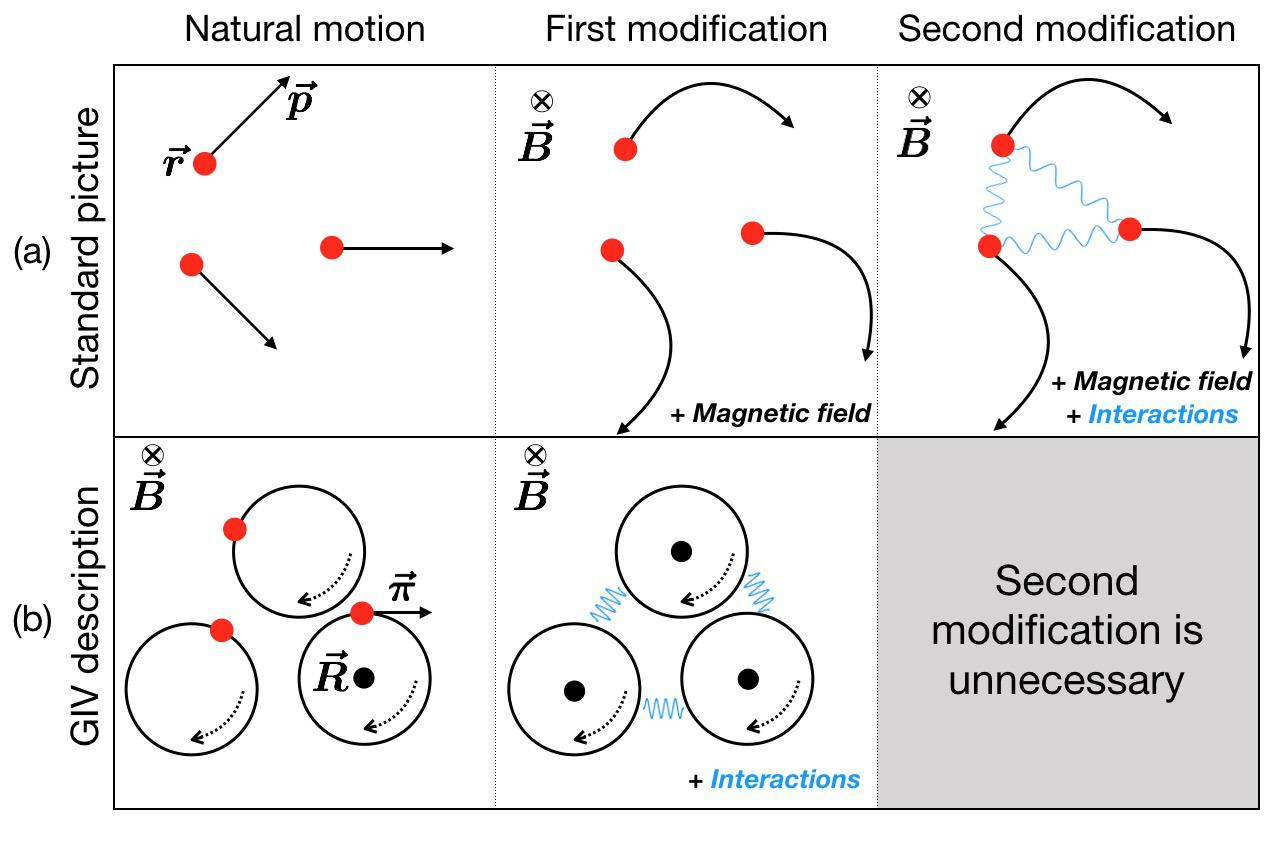}}%inclgphcs[trim=lcm bcm rcm tcm, clip=true, angle=-90]
\caption{Comparing kinematic formulations of planar quantum mechanics in the presence of a magnetic flux, $\mbB$, using (a) conventional position and linear momentum, vs.\ (b) gauge-invariant variables (GIVs), namely, the kinetic momenta, $\mbpp$, and the guiding center coordinates, $\mbR$. In (b), adding interactions is the only modification to the simple `free' picture necessary, for describing strongly correlated fractional quantum Hall phases. In contrast, in (a), first the magnetic field needs to be incorporated and then interactions are added in.}
\label{fig-GIVcompare}
\end{center}
\end{figure}
%%%%%%%%%%%%%%%%%%%%%%%end%%%%%%%%%%%%%%%%%%%%%%

In the absence of interactions, planar electrons in a perpendicular magnetic field can only have discrete energy values, corresponding to the well-known Landau levels \cite{1930-landau-fp}. Landau levels have macroscopic degeneracy, corresponding to the freedom of motion of the guiding center coordinates, which determine the spatial location of the electron's cyclotron orbit \cite{2002-yoshioka-zl,2007-jain-fk}. The presence of interactions splits this macroscopic degeneracy, giving rise to the FQH ground states at rational filling fractions. Since degenerate perturbation theory requires an exact many-body diagonalization of the interaction potential, finding the energy eigenstates corresponding to FQH ground and excited states, starting from the microscopic Hamiltonian, is a challenging problem. 

It is widely accepted that the exact form of the interaction potential does not affect the large-scale wavefunction structure of many FQH states, such as those at the simplest Laughlin fractions \cite{2007-jain-fk}. However, it is increasingly apparent that these details are important for determining the bulk structure of the technologically-relevant non-abelian FQH states (e.g., $\nu = 5/2$) \cite{1991-moore-uq,2007-lee-kk,2007-levin-ix,2013-balram-bq,2017-rezayi-cl,2017-mong-fq}.  They are also important for determining the details of edge reconstruction, critical for interpreting interferometric experiments probing the braiding statistics of quasiparticles \cite{2008-nayak-fk,2017-hansson-oq,2017-sabo-rq}.

In this work we use the method of coherent states \cite{1963-sudarshan-bs,1963-glauber-ve,1963-glauber-ek,2006-klauder-ox,2012-perelomov-kd} to show that the many-body guiding-center wavefunction, which encapsulates the physics of FQH behavior, has a one-to-one correspondence with complex multivariate holomorphic functions, with one complex variable per particle. Our result holds for states with \emph{any filling fraction}, irrespective of which Landau levels are involved. As a special case, our approach yields the holomorphic function representation of real-space wavefunctions in the lowest Landau level \cite{2002-yoshioka-zl,2007-jain-fk}. It also accounts for the success of similar representations for higher Landau level states, which arise from the use of conformal blocks \cite{1991-moore-uq,2017-hansson-oq}. Further, we derive the energy eigenvalue equation in this framework, Eq.~\eqref{eq-eigenMaster}, a qualitatively new form of Schr\"odinger's equation. Our formulation is apt for incorporating a new combination of techniques, from symmetric functions, Galois theory and complex analysis, to accurately predict the many-body energy eigenstates and eigenvalues corresponding to any FQH phase, starting from a microscopic Hamiltonian.

%\section{Gauge-Invariant Variables (GIV)}

\emph{Gauge-Invariant Variables (GIV).} Our approach utilizes Gauge-Invariant Variables (GIVs) \cite{1949-johnson-nr,2013-biswas-wr,2018-chen-zi}. To motivate their use, we contrast with the conventional kinematic description of two-dimensional motion of particles using real-space coordinates and linear momenta (see Figure~\ref{fig-GIVcompare}). The intuition underlying the conventional description is that free particles should move in straight lines. Upon introducing a perpendicular magnetic field, the Lorentz force causes these trajectories to bend into cyclotron orbits. Interactions add a second layer of complexity to the already-modified picture. Thus, the description of interacting particles in a magnetic field, namely, the physical situation where FQH states arise, is laborious in the conventional framework.

Our proposal to remedy this situation is to instead use a language which naturally includes the magnetic field in the `free' picture, thus leaving us to deal only with the addition of interactions to the problem. This is accomplished by using GIVs. In this framework, the `free' dynamical units are not the particles that move in straight lines, but rather, entire cyclotron orbits which are static in the absence of external fields (other than the background magnetic field). These orbits exhibit non-intuitive responses such as drifting perpendicularly to an in-plane electric field with a universal drift velocity $E/B$ \cite{2002-yoshioka-zl,1975-jackson-gl}. This universal drift velocity is fundamentally related to the quantization of Hall conductance.

In what follows we consider electrons (with charge $-e$) moving in an infinite flat plane. For this scenario, the GIVs are the well-known kinetic momenta and guiding center coordinates \cite{1949-johnson-nr}, respectively,
%%%%%%%%%%%%%%%%%%%%%%begin%%%%%%%%%%%%%%%%%%%%%%
\begin{align}\label{eq-GIV}
\mbpp = \mbp + e \mbA(\mbr), \quad \mbR = \mbr + \le(\ell^{2}/\hbar\ri)\hat{\mbz}\times\mbpp.
\end{align}
%%%%%%%%%%%%%%%%%%%%%%%end%%%%%%%%%%%%%%%%%%%%%%
$\ell = \sqrt{\hbar/eB}$ is the magnetic length and the magnetic field is $\mbB = B \hat{\mbz}$. These GIVs satisfy the commutation relations:
%%%%%%%%%%%%%%%%%%%%%%begin%%%%%%%%%%%%%%%%%%%%%%
\begin{align}\label{eq-GIVcommutation}
[R_{x}, R_{y}] = i \ell^{2}, \quad [\p_{y}, \p_{x}] = i \hbar^{2}/\ell^{2},
\end{align}
%%%%%%%%%%%%%%%%%%%%%%%end%%%%%%%%%%%%%%%%%%%%%%
while $\mbpp$ and $\mbR$ commute among themselves. The commutation relation of the guiding center coordinates captures the quantum geometry characterizing topological quantum systems. This is clear from analogous results in lattice systems \cite{2010-loring-jk} where this commutator has been related to the Chern number. 

For brevity, in what follows we have set the magnetic length ($\ell$), electronic charge ($e$) and $\hbar$ to unity. In these units, the commutation relations between the GIVs
are analogous to the canonical commutation relations between the $2D$ coordinates and canonical momenta:
%%%%%%%%%%%%%%%%%%%%%%begin%%%%%%%%%%%%%%%%%%%%%%
\begin{align}%\label{eq-}
[x, p_{x}] = i , \quad [y, p_{y}] = i,\quad [(x, p_{x}), (y, p_{y})] = 0.
\end{align}
%%%%%%%%%%%%%%%%%%%%%%%end%%%%%%%%%%%%%%%%%%%%%%
Thus, the GIVs can be obtained from canonical coordinates and momenta via a canonical transformation. Therefore, a unitary transformation relates the orthonormal quantum Hilbert space basis labeled by the coordinates, $\le\{\le|x,y\ri\ra\ri\}$, to another labeled by $\le\{\le|R_{x}, \p_{y}\ri\ra\ri\}$, the values of one operator from each of the canonical pairs in Eq.~\eqref{eq-GIVcommutation}. Consequently, the quantum wavefunction expressed in the GIV basis is a function of one component from each canonical pair in Eq.~\eqref{eq-GIVcommutation}, for e.g. $\psi(R_{x}, \p_{y})$. For more details see \cite{2013-biswas-wr,2018-chen-zi}. By straightforward generalization, the form of the many-body electronic wavefunction in the GIV language is $\Psi(\le\{R_{x}, \p_{y}\ri\}_{1}, \le\{R_{x}, \p_{y}\ri\}_{2} \ldots )$, where the numerical subscripts label particles. This wavefunction is completely antisymmetric under the permutation of every particle pair.

%\section{The Hamiltonian in GIV language}

\emph{The Hamiltonian in terms of GIVs.} We consider the following interacting electronic many-body Hamiltonian in a magnetic field:
%%%%%%%%%%%%%%%%%%%%%%begin%%%%%%%%%%%%%%%%%%%%%%
\begin{align}%\label{eq-}
\mc{H} = \sum_{\a} K(\mbp_{\a} + \mbA(\mbr_{\a})) + \sideset{}{'}\sum_{\a\b} U(|\mbr_{\a}-\mbr_{\b}|).
\end{align}
%%%%%%%%%%%%%%%%%%%%%%%end%%%%%%%%%%%%%%%%%%%%%%
The Greek indices label particles, and the prime on the second sum denotes a summation over distinct pairs. $K$ and $V$ are respectively the single-particle kinetic energy and the pairwise isotropic interaction potential.

The kinetic energy $K(\mbp + \mbA(\mbr))\equiv K(\mbpp)$ is a function only of the kinetic momenta and has a discrete spectrum. These discrete energies correspond to the well-known Landau levels, whose specification fixes the kinetic momentum part of the wave function. The exact form of $K$ and the presence of spin (or pseudospin) structure in the Hamiltonian do not alter our narrative. Therefore we ignore them in this presentation. In the absence of interactions, the single electron Hilbert space within a Landau level has macroscopic degeneracy, arising from the free guiding center part of the wavefunction, which does not affect the energy, since $\mbR$ commutes with $\mbpp$.

When the topmost Landau level is partially filled, weak interactions split the macroscopic Landau-level degeneracy and give rise to FQH physics. In this regime, we can renormalize the interaction by averaging over the fast motion of kinetic momenta:
%%%%%%%%%%%%%%%%%%%%%%begin%%%%%%%%%%%%%%%%%%%%%%
\begin{align}%\label{eq-}
\le\la U(|\mbr_{\a}-\mbr_{\b}|)\ri\ra_{\mbpp} &= \le\la U\le(\le|\mbR_{\a}-\mbR_{\b} + \hat{\mbz}\times(\mbpp_{\a}-\mbpp_{\b})\ri|\ri)\ri\ra_{\mbpp} \nn\\
&\equiv V(|\mbR_{\a}-\mbR_{\b}|).
\end{align}
%%%%%%%%%%%%%%%%%%%%%%%end%%%%%%%%%%%%%%%%%%%%%%
The renormalized interaction, $V$, depends on the Landau level structure and is a function only of the guiding center coordinates. This renormalization procedure involves \emph{all} Landau levels and incorporates inter-Landau-level correlations, critical for accurately capturing the physics at higher fillings \cite{1996-koulakov-dq}. (See Supplementary Section 1 and definition of generalized pseudopotentials below.) For the simple case when inter-Landau-level correlations are neglected, and only the physics of the topmost Landau level is considered, this reduces to the Landau level projection technique \cite{1990-duncan-le}. Additional techniques exist for incorporating the effects of Landau level mixing \cite{2013-biswas-wr,2018-chen-zi}, which we will explore elsewhere. Here we focus on the renormalized Hamiltonian:
%%%%%%%%%%%%%%%%%%%%%%begin%%%%%%%%%%%%%%%%%%%%%%
\begin{align}%\label{eq-}
\mc{H}_{\text{ren}} = \sum_{\a} K(\mbpp_{\a}) + \sideset{}{'}\sum_{\a\b} V(|\mbR_{\a}-\mbR_{\b}|).
\end{align}
%%%%%%%%%%%%%%%%%%%%%%%end%%%%%%%%%%%%%%%%%%%%%%
Since the kinetic and potential parts of this Hamiltonian commute, the energy eigenstates are products of Landau level kinetic momentum eigenfunctions of individual particles and a many-body wavefunction in guiding center space. When multiple Landau levels are involved, antisymmetrization entangles both GIV spaces in straightforward but complex ways, leading to interesting physics in states with filling fractions greater than one \cite{1983-halperin-tw,2007-jain-fk,2007-goerbig-gb}. 

For brevity, here we consider a single fractionally-filled Landau level and focus on the FQH physics induced by orbital interactions. We can show that the essential physics is unchanged when multiple Landau levels are occupied. (See Supplementary Section 1 for details.) The low-lying many-body energy eigenstates of such a partly filled Landau level are of the form
%%%%%%%%%%%%%%%%%%%%%%begin%%%%%%%%%%%%%%%%%%%%%%
\begin{align}%\label{eq-}
\Psi_{m}(\le\{\le(R_{x}, \p_{y}\ri)\ri\}) = \eta(\p_{y, 1})\eta(\p_{y, 2})\ldots \times\psi_{m}(\le\{R_{x}\ri\}).
\end{align}
%%%%%%%%%%%%%%%%%%%%%%%end%%%%%%%%%%%%%%%%%%%%%%
In this expression the braces denote the set of all particle coordinates, $\eta$ is the single particle Landau level kinetic momentum eigenfunction (it is a simple harmonic oscillator eigenfunction for isotropic dispersion \cite{2013-biswas-wr,2018-chen-zi}) and $\psi_{m}$ is a completely antisymmetric eigenfunction of the effective interaction:
%%%%%%%%%%%%%%%%%%%%%%begin%%%%%%%%%%%%%%%%%%%%%%
\begin{align}\label{eq-interactionEVequation}
\le[\sideset{}{'}\sum_{\a\b} V(|\mbR_{\a}-\mbR_{\b}|)\ri]\psi_{m} \equiv \mc{U}_{\text{eff}}\psi_{m} = \e_{m} \psi_{m}.
\end{align}
%%%%%%%%%%%%%%%%%%%%%%%end%%%%%%%%%%%%%%%%%%%%%%
It is straightforward to incorporate the non-interacting energy contributed by the kinetic part, $E_{K}$.  The many-body energies corresponding to $\Psi_{m}$ are simply:
%%%%%%%%%%%%%%%%%%%%%%begin%%%%%%%%%%%%%%%%%%%%%%
\begin{align}%\label{eq-}
E_{m} = E_{K} + \e_{m}.
\end{align}
%%%%%%%%%%%%%%%%%%%%%%%end%%%%%%%%%%%%%%%%%%%%%%
The set $\le\{\c_{m}(\{R_{x}\}), \e_{m}\ri\}$ encodes the FQH physics arising due to interactions. This form also explicitly demonstrates that FQH physics is of a $1+1$ dimensional nature, arising from guiding center dynamics \cite{2005-bergholtz-qe,2005-abanov-fj,2007-horsdal-yw}.

Finding solutions to Eq.~\eqref{eq-interactionEVequation} is a difficult problem. First, it is a many-body equation and hence a complex multivariate problem. The second impediment is that the two canonically conjugate variables, $R_{x}$ and $R_{y}$, appear with equal prominence in the operator $\mc{U}_{\text{eff}}$ in Eq.~\eqref{eq-interactionEVequation}. This feature is distinct from familiar physical problems, in which the canonical momentum and coordinate variables appear in distinct additive parts of the Hamiltonian, with different energy scales.  (A notable exception is the simple harmonic oscillator, which is satisfactorily solved only through an exact technique.) This feature of Eq.~\eqref{eq-interactionEVequation}  thus renders ineffective the usual approximation techniques for finding solutions. We propose to overcome this impediment by utilizing the language of coherent states, developed in the context of quantum optics \cite{1963-sudarshan-bs,1963-glauber-ve,1963-glauber-ek,2006-klauder-ox,2012-perelomov-kd}, wherein a similar challenge arises.

%\section{GIV coherent states}

\emph{GIV coherent states.} Consider indexing the single-particle one-dimensional guiding center Hilbert space by an orthonormal basis $\le\{\le|n\ri\ra\ri\}$, where $n$ is a non-negative integer.  Since the renormalized interaction potential, $V$, is rotationally invariant, we will choose the particular basis given by the eigenstates of $\mbR^{2}$: $\mbR^{2}\le|n\ri\ra = (2n + 1)\le|n\ri\ra$. These states are simple harmonic oscillator states in guiding center space. They are also eigenstates of the projected interaction, $V(\sqrt{2}|\mbR|)\le|n\ri\ra = V_{n}\le|n\ri\ra$. The $\le\{V_{n}\ri\}$ are \emph{generalized} pseudopotentials. (See Supplementary Section 1 for a general formula.) For the special case when only one Landau level is considered, they reduce to the standard Haldane pseudopotentials \cite{1990-duncan-le}. If $V$ has other symmetries, other choices for $\le\{\le|n\ri\ra\ri\}$ may be useful. Any quantum state in this Hilbert space can be uniquely expressed as a complex vector sum: $\le|\c\ri\ra = \sum_{n} \c_{n} \le|n\ri\ra$, with $\sum_{n}|\c_{n}|^{2} = 1$. The overcomplete basis of `coherent states', labelled by the complex variable $z$, is defined as follows \cite{1963-sudarshan-bs,1963-glauber-ve,1963-glauber-ek,2006-klauder-ox,2012-perelomov-kd}:
%%%%%%%%%%%%%%%%%%%%%%begin%%%%%%%%%%%%%%%%%%%%%%
\begin{align}%\label{eq-}
\le|z\ri\ra = e^{-|z|^{2}/2}\sum_{n} \f_{n}(z)\le|n\ri\ra, \quad \f_{n}(z) = \frac{z^{n}}{\sqrt{n!}}.
\end{align}
%%%%%%%%%%%%%%%%%%%%%%%end%%%%%%%%%%%%%%%%%%%%%%
The $\le\{\f_{n}\ri\}$ are \emph{holomorphic} functions whose choice above is motivated by the fact that we are considering motion in an infinite flat plane. $e^{-|z|^{2}/2}$ is a normalizing factor. Due to the orthonormality of the $\f_{n}$ states, $\iint_{\mb{C}} d^{2}z e^{-|z|^{2}} \f^{*}_{m}(z) \f_{n}(z) = \p \d_{mn}$, the coherent states satisfy the well-known completeness relation: $\iint_{\mb{C}} d^{2}z \le| z \ri\ra \le\la z \ri| = \p \mb{I}$. For other useful properties of coherent states we refer the reader to \cite{1963-sudarshan-bs,1963-glauber-ve,1963-glauber-ek,2006-klauder-ox,2012-perelomov-kd}. Using the definition of coherent states, we can map \emph{any} quantum state in guiding center space, $\le|\c\ri\ra = \sum_{n} \c_{n} \le|n\ri\ra$, to a unique holomorphic function:
%%%%%%%%%%%%%%%%%%%%%%begin%%%%%%%%%%%%%%%%%%%%%%
\begin{align}%\label{eq-}
\c(z) \equiv e^{|z|^{2}/2} \le\la \c | z\ri\ra = \sum_{n} \c_{n}^{*}\f_{n}(z).
\end{align}
%%%%%%%%%%%%%%%%%%%%%%%end%%%%%%%%%%%%%%%%%%%%%%

We have thus shown that the guiding center part of the wavefunction, whose properties are critical for uncovering FQH physics, can be described using holomorphic functions. This is true \emph{irrespective of the filling fraction and which Landau levels are occupied}. 

Our results have connections with the following known results. In the symmetric gauge, the quantum wavefunction in the lowest Landau level can be identified with a holomorphic function, $\c_{0}(z)$, where $z= x + iy$ (ignoring a fixed Gaussian factor)  \cite{2002-yoshioka-zl,2007-jain-fk}. This mathematical representation played a crucial role in identifying the Laughlin and related trial wavefunctions for FQH states in the lowest Landau level. An independent approach for generating real-space wavefunctions with desirable ground state characteristics involves using conformal blocks, which also give rise to approximately holomorphic functions \cite{1991-moore-uq,2017-hansson-oq}. Our analysis demonstrates that holomorphic functions can be used for describing states with multiple occupied Landau levels, due to the quantum geometry encapsulated by the commutation properties of the guiding center GIVs. We have also identified the precise universal relationship between these holomorphic functions and the microscopic many-body wavefunction in the coordinate basis. As a special case, in the lowest Landau level, our holomorphic function in the coherent state representation, $\c$, has a straightforward relation to the holomorphic function $\c_{0}$ characterizing the coordinate representation in the symmetric gauge: $\c(z) = \c_{0}(-i\sqrt{2}z^{*})^{*}$.

%\section{The GIV Schr\"odinger equation}

\emph{The GIV Schr\"odinger equation.} Thus far we have established that the many-body electronic wavefunction is a product of a set of non-interacting kinetic momenta wavefunctions (which correspond to fast high energy cyclotron orbit motion) and a many-body guiding center wavefunction (which captures the physics of FQH states). When multiple Landau levels and spins are involved, the permutation symmetries of the kinetic momentum and guiding center wavefunctions can be complicated, but are straightforward to incorporate \cite{1983-halperin-tw,2007-goerbig-gb}. For the simple case of spinless electrons in a single Landau level, the kinetic momentum wavefunction is symmetric and the guiding center wavefunction completely antisymmetric under particle permutation. The same simple symmetry structure is also applicable when multiple Landau levels are populated. (See Supplementary Section 1.) Once the permutation symmetry structure is fixed, the guiding center wavefunction corresponds to a many-variable holomorphic function $\c(\le\{z\ri\})$ with the same permutation properties.

We now present how the GIV holomorphic representation can be put to  practical use, by deriving the corresponding Schr\"odinger equation for determining the energy eigenstates. This is achieved by expressing Eq.~\eqref{eq-interactionEVequation} in the coherent state representation, $\la \c | \mc{U}_{\text{eff}} | \le\{z\ri\} \ra = \e \la \c | \le\{z\ri\} \ra = \e \c(\le\{z\ri\})$.  (Details are provided in the Supplementary Sections 2 and 3.) Briefly, the central challenge is to calculate matrix elements of the form $\la \c | V\le(|\mbR_{1}-\mbR_{2}|\ri) | \le\{z_{1}, z_{2}\ri\} \ra$. This is achieved by rotating from the commuting pair of canonical GIVs $(\mbR_{1},\mbR_{2})$ to another canonical commuting GIV pair: $(\mbR_{+},\mbR_{-})$, where $\mbR_{\pm} = (\mbR_{1} \pm \mbR_{2})/\sqrt{2}$. The interaction is a function only of $\mbR_{-}$ with eigenvalues that are the generalized pseudopotentials $\le\{V_{n}\ri\}$. (See Supplementary Section 1.) With this insight, we use well-known properties of coherent states to obtain the GIV Schr\"odinger equation:
%%%%%%%%%%%%%%%%%%%%%%begin%%%%%%%%%%%%%%%%%%%%%%
\begin{widetext}
\begin{align}\label{eq-eigenMaster}
\sum_{n=0}^{\infty} \frac{V_{n}}{n!} \sum_{i<j}  \le(\frac{z_{i} - z_{j}}{2}\ri)^{n} \le.\le(\frac{\pd}{\pd \x_{i}} - \frac{\pd}{\pd \x_{j}}\ri)^{n} \c\le(\le\{\x\ri\}\ri)\ri|_{\le\{\x_{k} = z_{k}, \x_{i} = \x_{j} = \frac{z_{i} + z_{j}}{2}\ri\}} = \e \c(\le\{z\ri\}).
\end{align}
\end{widetext}
%%%%%%%%%%%%%%%%%%%%%%%end%%%%%%%%%%%%%%%%%%%%%%
This equation is \emph{valid for any filling fraction}. It is derived directly from first principles and takes into account interactions between different Landau levels, i.e., filled Landau levels are not discarded as inert. To reiterate, $\c(\le\{z\ri\})$ describes electrons from \emph{all} occupied Landau levels.

Due to its novel form, the solution of the GIV Schr\"odinger equation, Eq.~\eqref{eq-eigenMaster}, requires the development of new techniques, to be presented in future works. As  previously noted, the symmetric gauge real space wavefunction in the lowest Landau level, $\c_{0}$, is related to $\c$ by the transformation $\c(z) = \c_{0}(-i\sqrt{2}z^{*})^{*}$. This transformation leaves Eq.~\eqref{eq-eigenMaster} intact; thus the same equation should also hold for real space wavefunctions in the lowest Landau level. In the context of coordinate basis wavefunctions, applicable only to the lowest Landau level, such an operator representation has appeared previously in \cite{1993-cappelli-fk}. The GIV Schr\"odinger equation, Eq.~\eqref{eq-eigenMaster}, is however applicable to any Landau level configuration and uses GIV wavefunctions.

To illustrate how Eq.~\eqref{eq-eigenMaster} can be utilized, we derive the following well known result: the Laughlin state at filling fraction $\nu=1/m$ is the exact unique ground state when the first $m-2$ pseudopotentials $(V_{1}, V_{2}\ldots V_{m-2})$ are positive and the rest are zero. (For details see Supplementary Section 4.) Briefly, this follows from the fact that each pairwise operator multiplying the pseudopotentials is a \emph{projection operator}. Thus, the eigenstates of a positive linear combination of these, the operator on the left side of Eq.~\eqref{eq-eigenMaster}, cannot have negative eigenvalues. At filling fraction $1/m$, when the first $m-2$ pseudopotentials are positive and the rest zero, the unique state with zero energy is the corresponding Laughlin state, thus proving that the Laughlin state must be the unique ground state.

%\section{Discussion}

\emph{Concluding remarks.} We have used the language of GIVs to derive a holomorphic representation of FQH physics, which is valid for \emph{any} Landau level filling pattern and for arbitrary forms of the kinetic energy. The framework that we have developed can be generalized to accomodate a variety of scenarios involving different real space manifolds and symmetries. We have demonstrated its use by formulating the FQH problem in this language, on an infinite plane with arbitrary isotropic pairwise interactions. Putting these insights and results together, we have derived the analyic GIV Schr\"odinger equation, Eq.~\eqref{eq-eigenMaster}, which naturally discards irrelevant high energy physics. The FQH many-body ground and excited state wavefunctions and energies correspond to the eigenstates and eigenvalues of this novel differential equation.

Eq.~\eqref{eq-eigenMaster} provides a new route for deriving the properties of FQH states from microscopic Hamiltonians, by recasting the quantum many-body calculation in the GIV representation. Since the wavefunction corresponds to a holomorphic function with certain permutation symmetries under particle exchange, our formulation also provides an avenue to exploit insights from diverse mathematical fields like symmetric polynomials, Galois theory, complex analysis, etc. We hope that a synthesis of our formalism with these techniques will allow for first-principles based predictions of FQH state properties, starting from realistic microscopic Hamiltonians. Apparently distinct descriptions of FQH physics, such as trial holomorphic wavefunctions in the lowest Landau level \cite{1983-laughlin-fk,2007-jain-fk}; the conformal block picture from conformal field theory \cite{1991-moore-uq,2017-hansson-oq}; the composite fermion approach \cite{2007-jain-fk}; topological quantum field theory \cite{2004-wen-gf}; and matrix product states \cite{2012-zaletel-vg} may be naturally unified in our coherent state GIV formulation.

\begin{acknowledgments}
Author contributions: This research was conceived of and designed by RRB. YK and RRB performed calculations and wrote the paper. We acknowledge useful discussions with Mike Manfra, Chris Greene, Gabor Csathy, Leonid Rokhinson and Srividya Iyer-Biswas. This research was supported by Purdue University Startup Funds and the Purdue Research Foundation.  This work was completed and presented to limited audiences before \cite{2018-haldane-er} was made available on arXiv.
\end{acknowledgments}

\end{document}